\documentclass[11pt,preprint]{aastex}

\slugcomment{to appear in the Astronomical Journal}

\begin{document}

\title{Radio continuum imaging of FIR luminous QSOs at $z>6$}

\author{C.L. Carilli, F. Walter}
\affil{National Radio Astronomy Observatory, PO Box O, Socorro, NM,
  USA 87801}
\email{ccarilli@nrao.edu}

\author{F. Bertoldi, K.M. Menten}
\affil{Max-Planck Instit\"ut f\"ur Radio Astronomie, Auf dem H\"ugel
  69, Bonn, Germany}

\author{X. Fan}
\affil{Steward Observatory, U. of Arizona, Tucson, AZ, USA}

\author{G.F. Lewis}
\affil{Dept. of Physics, University of Sydney, Sydney, Australia}

\author{Michael A. Strauss} 
\affil{Dept. of Astrophysical Sciences, Princeton U., Princeton, 
NJ, USA, 08544}

\author{Pierre Cox, A. Beelen}
\affil{Institu d'Astrophysique Spatial, U. de Paris-Sud, Orsay, 
France}

\author{A. Omont, N. Mohan}
\affil{Institute de Astrophysique, Paris, France}

\begin{abstract}

We present sensitive imaging at 1.4 GHz of the two highest redshift
far-infrared (FIR) luminous QSOs SDSS J114816.65+525150.2 ($z=6.42$)
and SDSS J104845.05+463718.3 ($z=6.2$). Radio continuum emission is
detected from J1148+5251 with $S_{1.4} = 55 \pm 12 \mu$Jy, while
J1048+4637 is marginally detected with $S_{1.4} = 26 \pm 12
\mu$Jy. Comparison of the radio and FIR luminosities shows that both
sources follow the radio-FIR correlation for star forming galaxies,
with implied (massive) star formation rates $\sim 10^3$ M$_\odot$
year$^{-1}$, although we cannot rule-out as much as 50$\%$ of the
FIR luminosity being powered by the AGN. 

Five bright ($> 22$ mJy) radio sources are detected within 8$'$ of
J1148+5251. This is a factor 30 more than expected for a random
field. Two sources have SDSS redshifts, including a $z = 1.633$ radio
loud quasar and a $z = 0.05$ radio galaxy.  However, we do not find
evidence for a galaxy cluster in the SDSS data, at least out to $z
= 0.2$. Considering the faint SDSS magnitudes of the remaining radio
sources, we conclude that the over-density of radio sources could
either be a statistical fluke, or a very large scale structure ($>
8$Mpc comoving) at $z \ge 1$.  We also consider the possibility of
gravitational lensing by the closest (in angle) bright galaxy in the
SDSS data at $z = 0.05$, and conclude that the galaxy provides
negligible magnification.

\end{abstract}

\keywords{radio continuum: galaxies --- infrared: galaxies ---
galaxies: active, starburst, formation, high redshift}

\object{J114816.65+525150.2, J104845.05+463718.3}

\section{Introduction}

We are pursuing an extensive study of the thermal dust, molecular
line, and radio continuum emission properties of luminous QSOs from
redshifts $z \sim 2$ into the epoch of reionization (EoR) at $z \ge
6$. One key aspect of these studies is to determine the star formation
characteristics of the host galaxies.  This question has become
paramount since the discovery of the bulge mass -- black hole mass
correlation in nearby galaxies, a result which suggests a fundamental
relationship between black hole and spheroidal galaxy formation
(Gebhardt et al. 2000; Ferarrese \& Merrit 2000).  We have found that
30$\%$ of optically selected QSOs are hyper-luminous far-infrared
(FIR) galaxies, with $\rm L_{FIR} > 10^{13}$ L$_\odot$, corresponding
to thermal emission from warm dust, and with dust masses $\ge 10^8$
M$_\odot$ (Omont et al. 2003; Carilli et al. 2001a; Bertoldi et
al. 2003a; Priddey et al. 2003). Radio continuum studies show that
most of these sources follow the radio-to-FIR correlation for star
forming galaxies (Carilli et al. 2001b).  If the dust is heated by
star formation, the implied (massive) star formation rates are of
order $10^3$ M$_\odot$ year$^{-1}$, enabling the formation of the
stellar content of a large spheroidal galaxy in a dynamical timescale
of order $10^8$ years, although a contribution to dust heating by the
AGN certainly cannot be ruled out (see Andreani et al. 2003). 
Demographic studies show that super-massive black holes
acquire most of their mass during major accretion events marked by the
QSO phenomenon (Yu \& Tremaine 2002).

Molecular line (CO) emission has been detected from 13 FIR-luminous $z
> 2$ QSOs to date (see Carilli et al. 2004 for a review), including
the highest redshift QSO, J1148+5251 at $z=6.42$ (Walter et al. 2003;
Bertoldi et al. 2003b).  The implied molecular gas masses are
$10^{10}$ to $10^{11}$ M$_\odot$.  Detecting large masses of dust and
CO in the highest redshift QSOs indicates that heavy element and dust
formation, presumably via star formation, is a fundamental aspect of
the early evolution of QSO host galaxies, right back to the EoR,
within 0.8 Gyr of the Big Bang. Indeed, these observations of the dust
and gas content of the highest redshift QSOs are currently the only
direct probe of the host galaxies in these systems. This conclusion is
supported by the super-solar quasar metallicities deduced from the
optical emission line ratios (Fan et al. 2003; Freudling et al. 2003;
Maiolino et al. 2003; Pentericci et al. 2002).

In this paper we present the most sensitive radio continuum
observations to date of the two highest redshift FIR-luminous QSOs
known, SDSS J114816.65+525150.2 at $z=6.42$ and SDSS
J104845.05+463718.3 at $z=6.2$ (hereafter J1148+5251 and
J1048+4637). Both sources have been detected at 250 GHz (1.2 mm) using
the MAMBO detector at the IRAM 30m telescope (corresponding to a
rest-frame wavelength of $160 \mu$m), with flux densities of
$5.0\pm0.6$ mJy and and $3.0\pm0.4$ mJy, respectively (Bertoldi et
al. 2003a). Sensitive, high resolution radio continuum observations,
in combination with the (sub)mm observations, potentially probe the
star formation characteristics of the host galaxies (Carilli et
al. 2001b). The radio observations presented herein were designed to
probe to levels expected for star forming galaxies, as set by the FIR
luminosities, as well as to perform wide-field imaging to study the
clustering of radio sources along the line of sight to the
QSOs. Clustering in the field is an important characteristic when
considering the possibility of gravitational magnification of the
QSO. Gravitational lensing is a key factor in the study of 
high-$z$ QSO demographics, and in calculations of 'cosmic Stromgren
spheres' (see discussion). We assume a standard concordance cosmology
with H$_o$ = 70 km s$^{-1}$ Mpc$^{-1}$.

\section{Observations and results}

Observations of J1148+5251 and J1048+4637 were made with the VLA at
1.4 GHz in the A and B configurations (30 km and 10 km maximum
baselines, respectively). The observational parameters are listed in
Table 1. Most of the observations were made in standard continuum
mode, except for one A array observation of J1148+5251 that was done
in spectral line mode in order to perform very wide field imaging.
Standard gain and phase calibration, and self-calibration, was
applied, and the images were made using the wide-field imaging
capabilities of the AIPS task IMAGR.

The last two columns in Table 1 give the measured flux density at the
optical position of the QSO for each observation (note: in all these
observations the sources are unresolved). For J1048+4637 we
get a (weighted) average value of $26 \pm 12\mu$Jy.  For J1148+5251
the value is $55 \pm 12\mu$Jy. Hence J1148+5251 is reasonably detected
at 4.6$\sigma$, while J1048+4637 is only marginally detected at
2.2$\sigma$. The B array images of the two sources are shown in
Figure 1 at 4.5$''$ resolution. 

Figure 2 shows the wide field image of J1148+5251, while Figure 3
shows a blow-up of the inner part of the field, along with the SDSS
optical image (the contour levels are different in the two images, in
order to emphasize only the brighter sources in Figure 2). Note the
large number of bright radio sources in the J1148+5251 field, in
particular five sources brighter than 22 mJy within 8$'$ of the
QSO. This is a factor 30 higher than the average source density for
$S_{1.4} > 22$mJy (White et al.  1997).  For comparison, the brightest
source within 8$'$ of J1048+4637 is 11 mJy, which is roughly as
expected for a random field.  Four of the five bright sources are
extended, double radio sources, and one is unresolved at 1.5$''$
resolution.  The SDSS reveals optical counterparts for all five
sources (York et al. 2000; Abazajian et al. 2003). Two of the sources
have spectroscopic redshifts, one corresponding to a bright QSO at $z
= 1.634$. The positions and magnitudes for the five sources brighter
than 22 mJy within $8'$ of J1148+5251 are given in Table 2, in order
of decreasing radio flux density.  We note that a similar over-density
of radio sources has also been found in the field around the $z=6.28$
QSO SDSS J1030+0524 (Petric et al. 2003).

The most striking radio source is the bright FRI (Fanaroff-Riley Class
I = 'edge-darkened') radio galaxy extending over 4$'$, with the core
located 1$'$ southwest of J1148+5251 (source 4 in Table 2).  This
galaxy was also detected at 250 GHz with $S_{250} = 4\pm 0.9$ mJy
using MAMBO (Bertoldi et al. 2003a), and  has an SDSS spectroscopic
redshift of $z=0.050$.  The SDSS photometry implies a bolometric
magnitude for this galaxy of M$_{bol} = -21.6$, or a bolometric
luminosity of $3.4\times 10^{10}$ L$_\odot$. There is also a companion
galaxy in the SDSS (not radio loud) at the same redshift, located
25$''$ northwest of the FRI host.

On seeing the over-density of bright radio sources in the J1148+5251
field, we searched the SDSS data for a galaxy cluster.  No cluster of
galaxies is seen within a degree of J1148+5251 out to $z = 0.2$,
although a linear 'filament' is detected in the galaxy distribution 
that extends from about 14,000 to 24,000 km s$^{-1}$. 

\section{Discussion}

\subsection{Star formation and the radio-FIR correlation}

The radio-FIR correlation for star forming galaxies is one of the
tightest correlations in extragalactic astronomy, and has become a
standard tool in the study of extragalactic star formation (Condon
1992), eg. as a fundamental assumption in the radio photometric
redshift technique, widely used for distant starburst galaxies
(Carilli \& Yun 1999), and as a radio-loud AGN diagnostic (Reddy \&
Yun 2004).  In this section we consider the two FIR-luminous $z>6$
QSOs in the context of the radio-FIR correlation.

The FIR luminosities of the sources were derived from the 250 GHz flux
densities by Bertoldi et al. (2003a) assuming a spectral energy
distribution (SED) typical of an ultra-luminous infrared galaxy,
corresponding roughly to thermal emission from warm dust at 50~K with
a dust emissivity index $\beta = 1.5$.  For J1148+5251 they find: $\rm
L_{FIR} = 1.21\pm0.14 \times 10^{13}$ L$_\odot$.  The value for
J1048+4637 is $\rm L_{FIR} = 7.5 \pm 1.0 \times 10^{12}$ L$_\odot$.

We derive the intrinsic luminosity density at 1.4 GHz from the
observed flux density at 1.4 GHz assuming a spectral index of $-0.8$,
characteristic of star forming galaxies.  For J1148+5251 we find:
$L_{1.4} = 2.1\pm0.4 \times 10^{25}$ W Hz$^{-1}$, and for J1048+4637:
$L_{1.4} = 9.2\pm4.6 \times 10^{24}$ W Hz$^{-1}$. 

The radio-FIR correlation has been quantified via the standard $q$
parameter, defined as (Condon 1992): 

$$q \equiv \rm log({{L_{FIR}}\over{3.75~\times~10^{12}~W}}) 
- log({L_{1.4}\over{W ~ Hz^{-1}}})$$

\noindent where L$_{\rm FIR}$ is defined as the far infrared
luminosity between 40 and 120 $\mu$m.  For J1148+5251 we find: $q =
1.80\pm0.14$, and for J1048+4637: $q = 1.89\pm 0.36$, where the errors
are derived assuming $\pm 1 \sigma$ errors on each quantity.

In Figure 4 we plot the $q$ values versus 60$\mu$m
luminosity\footnote{ For the assumed SED the 60$\mu$m luminosity ($\rm
\nu L_\nu$) is 2/3 the FIR luminosity.}  for J1148+5251 and
J1048+4637, and for the IRAS 2 Jy sample of galaxies from Yun et
al. (2001).  The solid line shows the mean of the relationship for the
IRAS galaxies, $q = 2.35$, while the dashed lines show the range
determined by Yun et al. to correspond to star forming galaxies.
Objects below this range are radio-loud AGN.  Both the high $z$ QSOs
fall within the range corresponding to star forming galaxies. If star
formation is the dominant dust heating mechanism in these sources, the
implied star formation rates are about 3000 M$_\odot$ year$^{-1}$ and
2000 M$_\odot$ year$^{-1}$, respectively (Bertoldi et al. 2003b).

Of course, we cannot rule-out some contribution to dust heating by the
AGN. For instance, the sources could still fall in the range defined
by star forming galaxies (ie. $q > 1.6$) and yet have $30\%$ to $50\%$
of their FIR luminosity powered by AGN dust heating.  Likewise,
current constraints on the rest-frame IR SEDs cannot preclude a warm
($> 100$K) dust component, presumably heated by the AGN, that would
dominate in the mid-IR, as has been seen in the cloverleaf QSO at
$z=2.558$ (Weiss et al. 2003).

The radio-FIR correlation provides a consistency check for star
formation in the host galaxies of these systems, although it should
not be construed as proof thereof.  Coupled with the large molecular
gas masses, the fuel for star formation, these observations argue for
a massive starburst coeval with the accretion activity onto the
supermassive black hole in these systems.  High resolution imaging of
the CO emission is in progress which should help elucidate the
characteristics of the host galaxy (Walter et al. 2004).

\subsection{Gravitational Lensing of J1148+5251}

Strong gravitational lensing would magnify these sources, such that
the inferred luminosities would be over-estimated.  The intrinsic
luminosity is one of the key parameters in the calculation of the
radii of cosmic 'Stromgren spheres' around these sources, which can be
used to constrain the timescale for QSO activity, or alternatively,
the neutral fraction of the IGM (Haiman \& Cen 2002; White et
al. 2003; Wyithe \& Loeb 2004).  Lensing corrections are also
fundamental to the derivation of QSO demographics (Wyithe \& Loeb
2002; Fan et al. 2003; Yu \& Tremaine 2002; Richards et al. 2004).

Although there is a clear over-density of bright radio sources in the
J1148+5251 field, no galaxy cluster is seen in the SDSS data out to $z
= 0.2$.  Two of the radio sources have spectroscopic redshifts ($z =
0.05$ and $z=1.6$). If we assume that the radio sources without
spectroscopic redshifts have host galaxies with luminosities similar
to that of the $z = 0.05$ source, then their faint optical magnitudes
place them at $z \sim 1$ to 1.5. Hence, it is possible that these
three sources, plus the QSO, form a large scale structure at $z \sim
1.6$.  However, the angular separation of the sources implies a 
scale for such a structure $\ge 8$ Mpc (comoving), ie. not a dense
cluster.  We conclude that the over-density of bright radio sources in
the J1148+5251 field is either a statistical fluke, or a very large
scale structure possibly at $z \sim 1.6$. Such filamentary structures
on very large scale lack the mass surface density to produce
strong gravitational magnification (Bacon et al. 2001).

We also consider lensing by the brightest galaxy closest to the
line-of-sight to the QSO at $z = 0.05$ (source 4 in Table 2).  From
the observed magnitude we calculate a velocity dispersion of 275 km
s$^{-1}$ for this galaxy using the Faber-Jackson relation. The
magnification of the QSO 1$'$ distant from the galaxy is then
negligible (roughly $5\%$).

Overall, the radio and SDSS data provide no evidence for a dense
cluster, or bright galaxy, along the line of sight to J1148+5251 which
could act to magnify the QSO. Of course, these data cannot rule-out a
much more distant galaxy, or cluster, close to the line of sight
lensing the QSO.  White et al. (2003) consider this question in
detail, and in particular, they discuss evidence for a possible
'proto-cluster' at $z\sim 5$, as suggested by Ly $\alpha$ emission and
CIV absorption in the QSO spectrum.  They conclude that such a
distant cluster could magnify substantially the QSO without violating
the size constraint of $0.4"$ set by ground-based near-IR imaging (Fan
et al. 2003). HST imaging at $0.1"$ is in progress which will test
this interesting possibility (White et al. in prep).  We note that HST
images of four other $z > 5.7$ QSOs by Richards et al. (2004) show no
evidence for multiple imaging on scales of 0.1$"$.

\acknowledgments

The National Radio Astronomy Observatory is operated by Associated
Univ. Inc., under contract with the National Science Foundation.

\clearpage


\begin{deluxetable}{cccccccccc}
\tabletypesize{\scriptsize}
\tablecaption{Observational Parameters \label{tbl-1}}
\tablewidth{0pt}
\tablehead{
\colhead{Source} & \colhead{Date} & \colhead{Config.} &
\colhead{Mode} & \colhead{Bandwidth} & \colhead{Time} & 
\colhead{Frequency} & \colhead{FWHM} & \colhead{$S_{1.4}$} &
\colhead{rms}} 
\startdata
~ & ~ & ~ & ~ & MHz & hours & MHz & arcsec & $\mu$Jy & $\mu$Jy \\
\hline 
J1048+4637 & June 20, 2003 & A & Continuum & $2\times50$ MHz & 2
& 1385, 1465 & 1.5 & 28 & 19  \\
J1048+4637 & Jan 17, 23, 2004 & B & Continuum & $2\times50$ MHz
& $2\times5$ & 1385, 1465 & 4.5 & 26 & 13 \\
J1148+5251 & June 20, 2003 & A & Continuum & $2\times50$ MHz & 2
& 1385, 1465 & 1.5 & 46 & 20 \\
J1148+5251 & Sept 19, 2003 & A & Line, 8chan & $2\times25$ MHz & 8
& 1365, 1435 & 1.5 & 57 & 18 \\
J1148+5251 & Jan 16, 2004 & B & Continuum & $2\times50$ MHz
& 4 & 1385, 1465 & 4.5 & 67 & 27 \\
\enddata
\end{deluxetable}

\begin{deluxetable}{ccccc}
\tabletypesize{\scriptsize}
\tablecaption{Radio sources brighter than 22 mJy within $8'$ of 
J1148+5251 \label{tbl-2}} 
\tablewidth{0pt}
\tablehead{\colhead{~} & 
\colhead{Position (J2000)} & \colhead{r$^a$} & \colhead{$S_{1.4}$} 
& \colhead{z}}
\startdata
~ & h m s, ~~ d m s & mag & mJy & ~ \\
\hline 
1 & 11 48 56.57, +52 54 25.3 & 16.57 & 100 & 1.633 \\
2 & 11 48 16.10, +52 58 59.2 & 20.38 & 98 &  ~ \\
3 & 11 48 19.59, +52 52 12.3 & 20.62 & 84 &  ~ \\
4 & 11 48 12.19, +52 51 07.5 & 14.70 & 56 &  0.050 \\
5 & 11 47 59.11, +52 55 42.8 & 20.42 & 22 &  ~ \\
\enddata
\vskip 0.1in
~$^a$Model magnitudes from the Sloan Digital Sky Survey
(York et al. 2000; Abazajian et al. 2003).
\end{deluxetable}

\clearpage\newpage

\begin{figure}
\plotone{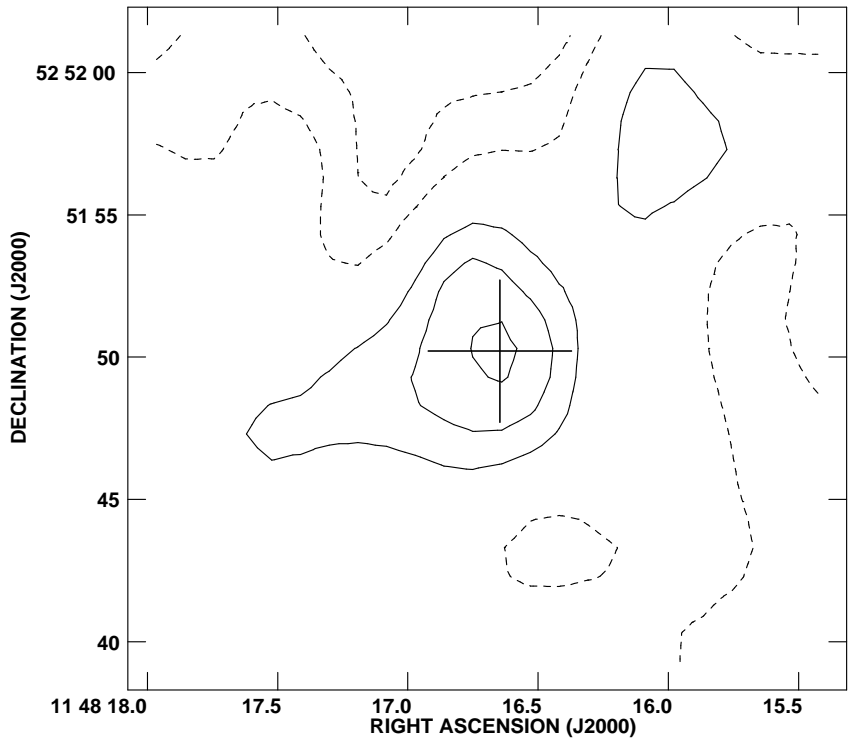}
\vspace*{-1in}
\caption{VLA image of the SDSS J1148+5251 at 1.4 GHz at
4.6$''$ resolution (FWHM). The contour levels are --52, --26, 26, 52,
78 $\mu$Jy beam$^{-1}$. The QSO position is indicated by a cross.}
\end{figure}

\clearpage

\begin{figure}
\plotone{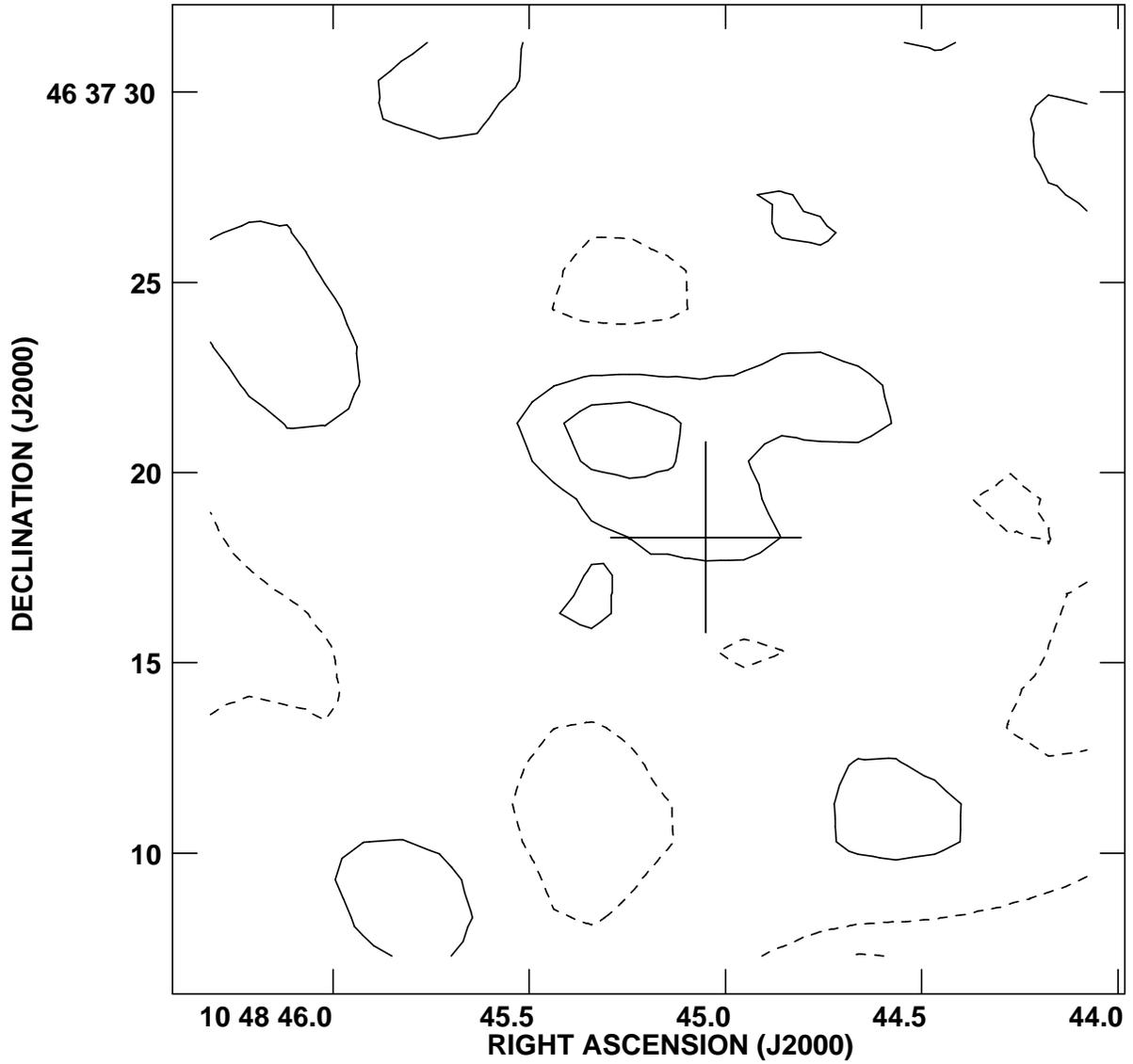}
\caption{VLA image of the SDSS J1048+4637 at 1.4 GHz at
4.6$''$ resolution (FWHM). The contour levels are --36, --18, 18, 36, 54
$\mu$Jy beam$^{-1}$. The QSO position is indicated by a cross.}
\end{figure}

\clearpage

\begin{figure}
\plotone{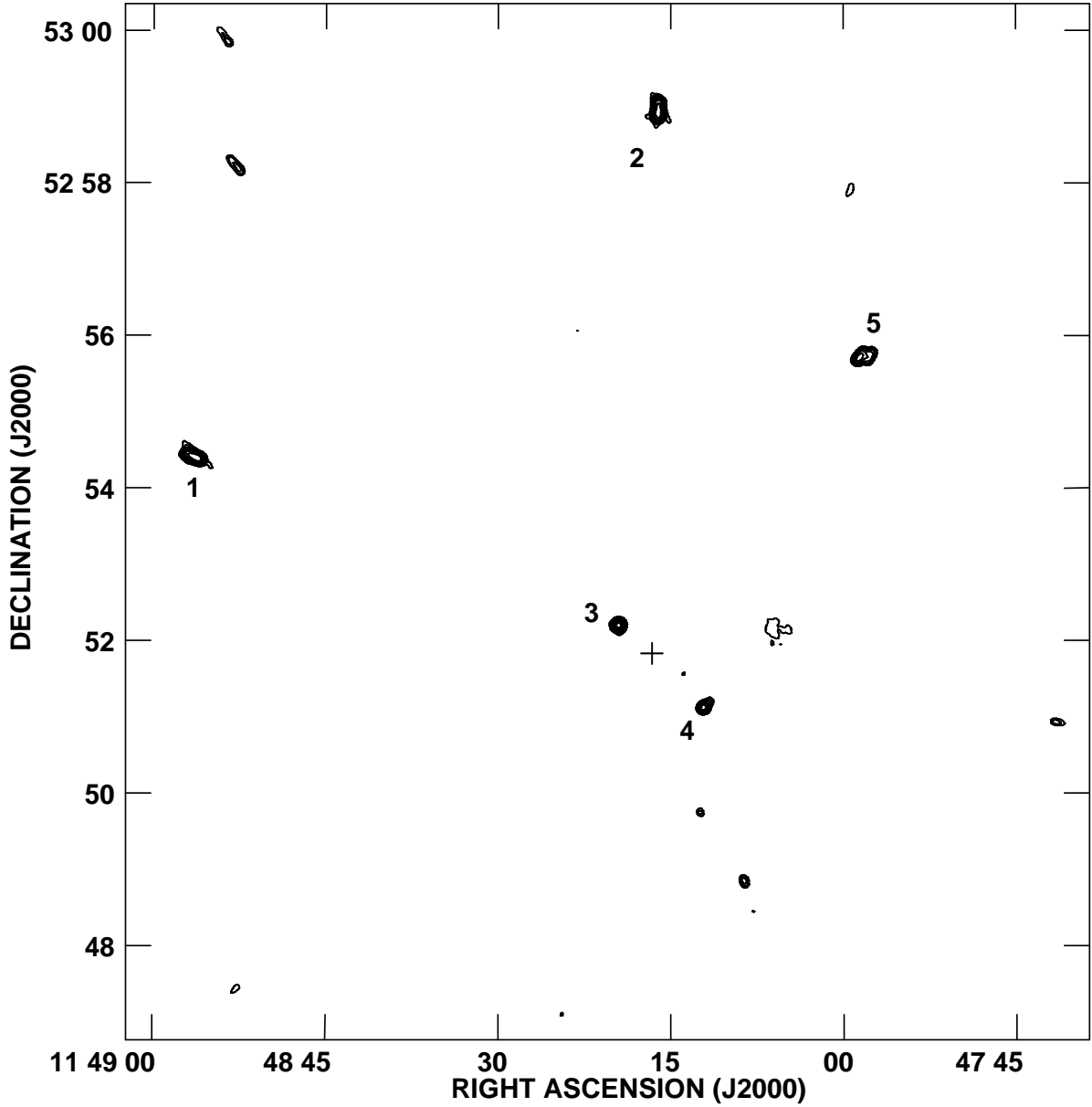}
\caption{VLA image of the SDSS J1148+5251 field at 1.4 GHz at
4.6$''$ resolution (FWHM). The contour levels are a geometric
progression by a factor of two, starting at 0.2 mJy beam$^{-1}$. 
The QSO position is indicated by a cross.  The elongation of the
outer-most sources ($\ge 6'$ from the QSO) is due to 
bandwidth smearing.}
\end{figure}

\clearpage

\begin{figure}
\plotone{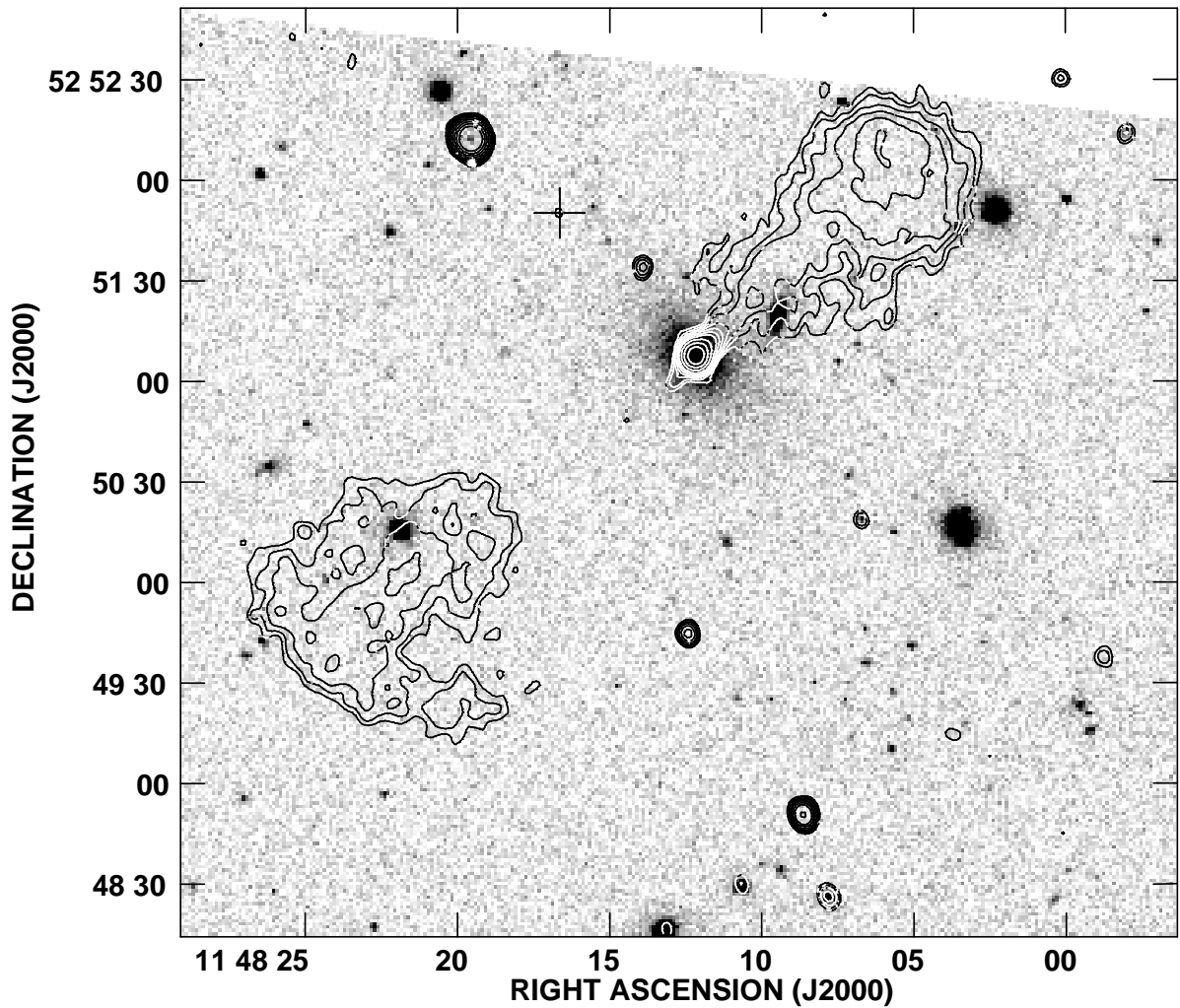}
\caption{VLA image of the inner part of the 
SDSS J1148+5251 field at 1.4 GHz at
4.6$''$ resolution (FWHM). The contour levels are a geometric
progression by a factor of $\sqrt{2}$, starting at 0.075 mJy 
beam$^{-1}$.  The QSO position is indicated by a cross. Note
that the contour levels go to lower levels than in Figure 1, 
in order to emphasize the extended structures of the radio
galaxy 1$'$ southwest of the QSO. 
The grayscale is the SDSS image (York et al. 2000; Azabajian et 
al.  2003).}
\end{figure}

\clearpage

\begin{figure}
\plotone{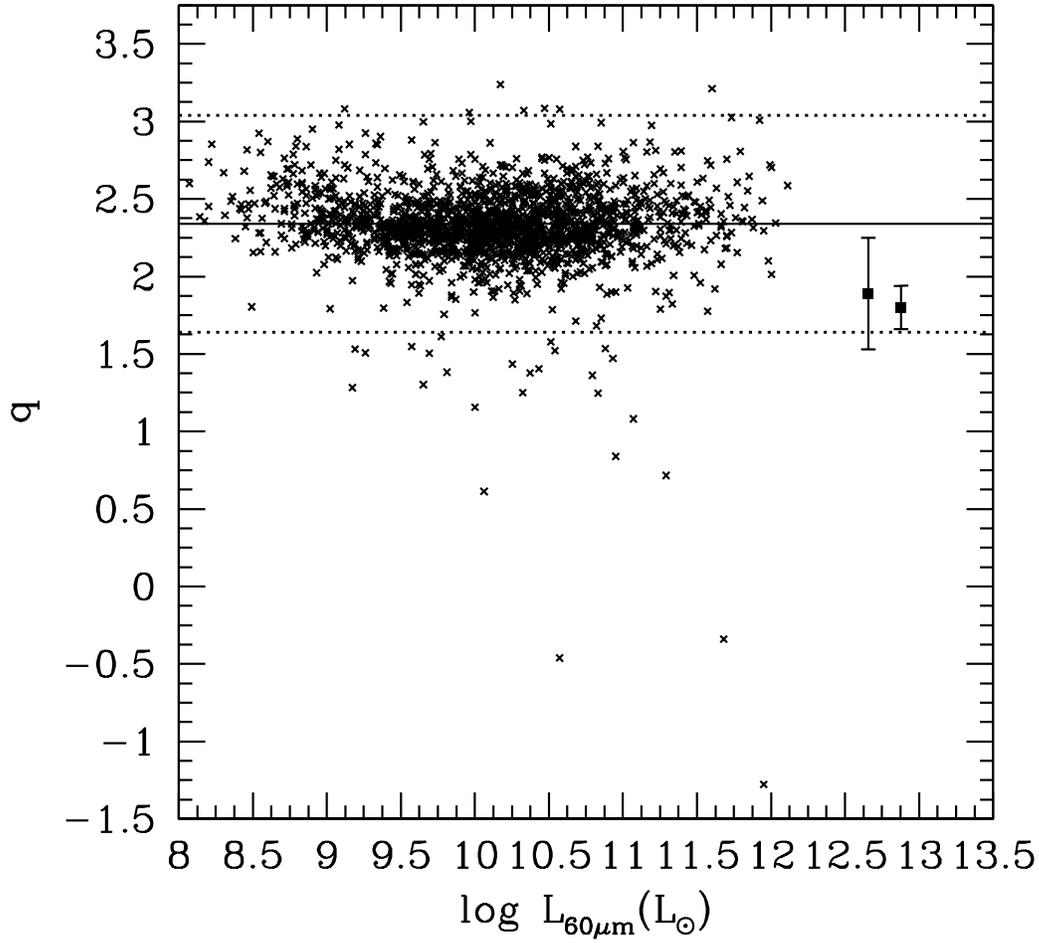}
\caption{The radio-FIR correlation for the IRAS 2Jy galaxy sample
(x) from Yun et al. (2001), and for J1148+5251 and J1048+4637
(points with error bars).  The $q$ parameter and L$_{\rm 60\mu m}$ 
are  defined in section 3.1. The solid line is the mean value
for the IRAS galaxies, and the dotted lines indicate the range
defined as star forming galaxies by Yun et al. (2001).}
\end{figure}

\clearpage\newpage

\end{document}